\documentclass[11pt,fleqn,twoside]{article}
\usepackage{emlines,latexsym}
\makeatletter
\newcommand{\prava}{\footnotesize\it
\begin{flushright}
\begin{minipage}{6cm}%9.6
Copyright \copyright 1998 by M.B. Abd-el-Malek and M.N. Makar
\end{minipage}
\end{flushright}}

\newcommand{\name}[1]{\begin{flushleft}
                       \LARGE \bf #1
                       \end{flushleft}\vspace{-3mm}}

\newcommand{\Author}[1]{\begin{flushleft}
                       \it #1 \end{flushleft}}

\newcommand{\Adress}[1]{\begin{flushleft}
                       \it #1 \end{flushleft}}

\newcommand{\Date}[1]{\begin{flushleft}
                      \small  \it #1 \end{flushleft}}

\newcommand{\ehkol}{Author \ name}
\newcommand{\ohkol}{Article \ name}
\renewcommand{\@evenhead}{
\hspace*{-3pt}\raisebox{-15pt}[\headheight][0pt]{\vbox{\hbox to \textwidth
{\thepage \hfil \ehkol}\vskip4pt \hrule}}}
\renewcommand{\@oddhead}{
\hspace*{-3pt}\raisebox{-15pt}[\headheight][0pt]{\vbox{\hbox to \textwidth
{\ohkol \hfil \thepage}\vskip4pt\hrule}}}
\renewcommand{\@evenfoot}{}
\renewcommand{\@oddfoot}{}

\newcommand{\be}{\begin{equation}}
\newcommand{\ee}{\end{equation}}
\newcommand{\ba}{\hspace*{-5pt}\begin{array}}
\newcommand{\ea}{\end{array}}
\newcommand{\p}{\partial}
\newcommand{\ds}{\displaystyle}
\makeatother

\begin{document}
\setcounter{page}{41}

\thispagestyle{empty}

\renewcommand{\ehkol}{M.B. Abd-el-Malek  and M.N. Makar}
\renewcommand{\ohkol}{Progressive Internal Gravity Waves}

\begin{flushleft}
\footnotesize \sf
Journal of Nonlinear Mathematical Physics \qquad 1998, V.5, N~1,\ 
\pageref{malek-fp}--\pageref{malek-lp}.\hfill {\sc Article}
\end{flushleft}

\vspace{-5mm}

{\renewcommand{\footnoterule}{}
{\renewcommand{\thefootnote}{}  \footnote{\prava}}

\name{Progressive Internal Gravity Waves With Bounded Upper Surface
Climbing a Triangular Obstacle}\label{malek-fp}

\Author{Mina B. ABD-EL-MALEK~$\dag$  and Malak N. MAKAR~$\ddag$}

\Adress{$\dag$~Department of Engineering Mathematics, Faculty of
Engineering,\\
~~Alexandria University, Alexandria 21544, Egypt\\
~~E-mail: mina@alex.eun.eg
\\[2mm]
$\ddag$~Department of Mathematics, Faculty of Science,\\
~~Assiut University, Assiut 71516, Egypt}

\Date{Received January 10, 1997; Accepted July 10, 1997}

\begin{abstract}
\noindent
In this paper we discuss a theoretical model for the interfacial
prof\/iles of progressive non-linear waves which result from introducing a
triangular obstacle, of f\/inite height, attached to the bottom below
the f\/low of a stratif\/ied, ideal, two layer f\/luid, bounded from above
by a rigid boundary. The derived equations are solved by using a
nonlinear perturbation method. The dependence of the interfacial
prof\/ile on the triangular obstacle size, as well as its dependence on
some f\/low parameters, such as the ratios of depths and densities of
the two f\/luids, have been studied.
\end{abstract}

\section{Introduction}
The determination of f\/low patterns over obstacles is a problem of a
great interest, that attracted many scientists over the past decades.
Lamb [6] was the f\/irst to give the essential features of the f\/low of
an ideal f\/luid in an open channel in the presence of an obstruction
in the channel. In 1955, Long [7] and then later on McIntyre [8] in
1972, considered the case of a steady and uniform stratif\/ication over
obstacles of f\/inite height, while Mei and LeMehaute [9] in 1966
studied the case of long waves in shallow water over an uneven
bottom. The ef\/fect of the irregularities of the bottom, on gravity
waves, has been studied by Kakutani [4] in 1971 via a reductive
perturbation method. Recently, Kevorkian and Yu [5], in 1989, studied
the behaviour of shallow water waves excited by a small amplitude
bottom disturbance in the presence of a uniform incoming f\/low. In
this paper we study a theoretical model for the interfacial prof\/iles
of progressive non-linear waves result from introducing a triangular
obstacle, attached to the bottom below the f\/low of a stratif\/ied,
ideal, two layer f\/luid, bounded from above by a rigid boundary. Our
primary motivation for the present investigation is calculate the
shape of the interfacial waves, and to discuss the inf\/luence of both
geometrical and f\/low parameters on the prof\/iles. In section 2, we
extended the mathematical technique applied by Helal \& Molines [3],
1981, in determining the nonlinear free-surface and
interfacial waves in a tank with f\/lat
horizontal bottom. Nonlinear perturbation technique is used,
leading, in sections 3 and 4, to expression for interfacial wave, was
derived in the form of expansions in powers of
$\varepsilon^2$, where $\varepsilon$ is a small parameter that provides a
measure of weakness of dispersion. Boutros, {\it et al} [1], in 1991,
applied the same technique to study the internal waves over a ramp.

Finally, in section 5 we have
discussed the ef\/fect of the density ratio, $R$, the thickness ratio, $H$,
and the triangle height, $L$.

}

\section{Formulation of the problem}

Two-dimensional irrotational
motion of a stably stratif\/ied two-layer f\/luid with a rigid upper
boundary and a bottom surface in the form of a triangle with two
inclination angles $\alpha$ and $\beta$. A cartesian coordinate system is
def\/ined with the origin at the bottom surface.

We assume that the
motion is two-dimensional, and the f\/luid is inviscid, incompressible,
and that the f\/low f\/ield due to the wave motion remains irrotational
and consequently we can introduce velocity potentials of upper and
lower layers are denoted by $\Phi^{*(i)}$, $i=1,2$,
respectively. Moreover, let $H_i^*$, $\rho^{(i)}$;
$i=1,2$  denote the thickness and densities of the upper
and lower f\/luids, respectively, $\tau^*$ is the time,
$Y^*=W^*(X^*)$ is the bed of the
channel and $Y^*=f^*(X^*, \tau^*)$ is the
interfacial disturbance from uniform condition. The component of
gravity, vertically downwards, is $g$. The equations of motion are thus
the Euler equations together with the continuity equation. All
variables are non-dimensionalized by using the
characteristic length $H_2^*$ and time
$(g/H_2^*)^{-1/2}$, and accordingly
\[
U=U^*/[gH_2^*]^{1/2} \qquad \mbox{and} \quad \Phi^{(i)}
=\Phi^{*(i)}/\left(H_2^*[gH_2^*]^{1/2}\right).
\]
Moreover, assuming that the f\/luids are in the
undisturbed uniform state up/down stream at inf\/inity, we
impose the following boundary conditions with respect to $X^*$
\[
\Phi_{X^*}^{*(i)}=U^*, \qquad (i=1,2) \qquad \mbox{as} \quad X^*\to
\pm \infty.
\]
An essential step which makes our
problem easier in handling is to def\/ine an appropriate
{\it stretching} of the horizontal coordinate
while leaving the vertical coordinate unchanged due to the fact that
the horizontal dimensions are much greater than the vertical
dimensions, thus we def\/ine
\be  \label{2.3}
x=\varepsilon X, \qquad y=Y, \qquad t=\varepsilon\tau,
\ee
where $\varepsilon$ is a small parameter.
Thus the basic equations for this system can be written as
\be  \label{2.4.1}
\varepsilon^2\Phi_{xx}^{(1)} +\Phi_{yy}^{(1)}=0, \qquad f<y<1+H,
\quad -\infty<x<\infty,
\ee
\be  \label{2.4.2}
\varepsilon^2\Phi_{xx}^{(2)} +\Phi_{yy}^{(2)}=0, \qquad W<y<f,
\quad -\infty<x<\infty,
\ee
with conditions\\
(i) Boundary conditions:
\be  \label{2.4.3}
\left.
\ba{l}
\Phi_y^{(i)}=\varepsilon f_t+\varepsilon^2\Phi_x^{(i)}f_x, \qquad
(i=1,2)\\[3mm]
\ds R\left\{\varepsilon \Phi_t^{(1)}+\frac 12 \left[ \varepsilon^2
(\Phi_x^{(1)})^2 + (\Phi_y^{(1)})^2\right] +f-1\right\}=\\[3mm]
\ds \left\{\varepsilon \Phi_t^{(2)}+\frac 12 \left[ \varepsilon^2
(\Phi_x^{(2)})^2 + (\Phi_y^{(2)})^2\right] +f-1\right\}
\ea
\right\} \qquad \mbox{at} \quad y=f
\ee
\be  \label{2.4.4}
\Phi_y^{(2)}=\varepsilon^2\Phi_x^{(2)}W_x \qquad \mbox{at} \quad
y=W(x)
\ee
\be    \label{2.4.5}
\Phi_y^{(i)}=0, \qquad \mbox{at} \quad  y=1+H,
\ee
\be  \label{2.4.6}
\varepsilon\Phi_x^{(i)}=1, \qquad (i=1,2) \quad \mbox{as} \quad x\to
\pm\infty
\ee
(ii) Initial condition:

at $t=0$: the initial prof\/ile of the interfacial wave, denoted by
$f(x,0)$, is shown in Fig.1.

\vspace{7mm}

{
\footnotesize

\hspace*{2cm}\input{malek.pic}

\vspace{-3mm}

\noindent
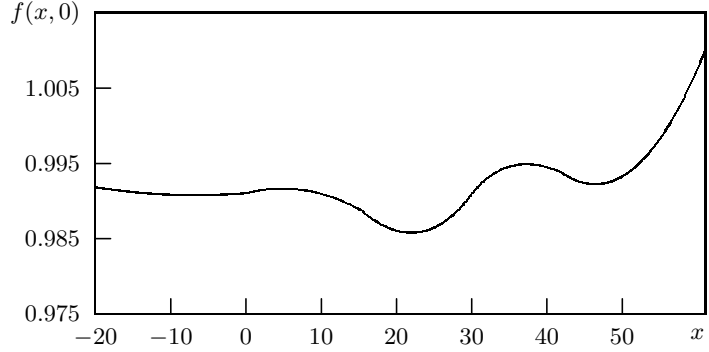
Fig.1. The inital waveform over a triangular obstacle with $x_m=13$,
$x_e=30$, $L=0.25$, $\alpha=0.01923$, $\beta=-0.01471$}

\vspace{5mm}

\noindent
where the density ratio $R= \rho^{(1)}/\rho^{(2)}$ (less than unity) and the
thickness ratio $H$ are two characteristic parameters of the system,
and $W(x)$ has the form
\be    \label{2.5}
W(x) = ax + b,
\ee
where
\[
(a,b)=\left\{
\ba{ll}
(0,0) & x\leq 0\\
(\alpha,0) & 0\leq x\leq x_m\\
(-\beta, \beta x_e) & x_m\leq x\leq x_e\\
(0,0) & x>x_e
\ea
\right.
\]
Since we consider weakly nonlinear waves, we expand the dependent
variables as power series in the same parameter $\varepsilon$ around the
undisturbed uniform state, following Helal and Molines [3], we get
\[
\ba{l}
\ds\Phi^{(i)}=\sum\limits_{n=0}^\infty \varepsilon^{2n-1}
G_{2n-1}^{(i)}(x,y,t), \qquad i=1,2\\[3mm]
\ds f=\sum\limits_{n=0}^\infty \varepsilon^{2n} f_{2n}(x,y,t),
\qquad \mbox{with} \quad f_0=1.
\ea
\]

The scale parameter $\varepsilon$, which is assumed to be small, provides a
measure of weakness of dispersion.

The boundary conditions on the
interface, equations (\ref{2.4.3}), are expanded as a Taylor
expansion of the type
\be \label{2.8}
[V]_{y=y_0+\varepsilon^2A} =\sum_{n=0}^\infty \frac{(\varepsilon^2
A)^n}{n!} \left[ \frac{\p^n V}{\p y^n}\right]_{y_0}.
\ee
When (\ref{2.3}), (\ref{2.5}), using the
expansion (\ref{2.8}), are inserted into equations (\ref{2.4.1})-(\ref{2.4.6})
and powers of $\varepsilon$ are sorted out, we get an ordered set of equations
to be solved.

\section{Orders of approximations}

\subsection{The f\/irst-order approximation}

Equations of the f\/irst-order approximation, f\/inally gives, for
$i=1,2$
\[
G_1^{(i)}= B^{(i)}(x,t),
\]
where $B^{(i)}(x,t)$ are unknown functions to be determined.

\subsection{The second-order approximation}

From the equations obtained from the second-order approximation, we
conclude that
\[
B_x^{(i)}=0, \qquad (i=1,2) \quad \mbox{as} \quad x\to \pm\infty
\]
and
\[
f_2(x,t)=\frac{1}{1-R} \left[ RB_t^{(1)}-B_t^{(2)}\right].
\]

\subsection{The third- and fourth-order appoximations}

Equations of the third- and fourth-order approximation, f\/inally
gives, for $i=1,2$
\be \label{3.3.1}
G_3^{(i)}=-\frac 12 y^2 B_{xx}^{(i)} +yC^{(i)}(x,t)+D^{(i)}(x,t),
\ee
where $C^{(i)}(x,t)$ and $D^{(i)}(x,t)$  are arbitrary functions
satisfy the following boundary conditions:
\be \label{3.3.2}
C_x^{(i)}=0 \qquad (i=1,2) \quad \mbox{as} \quad x\to \pm\infty,
\ee
\be   \label{3.3.3}
C^{(2)}(x,t)=\left(WB_x^{(2)}\right)_x \qquad \mbox{at} \quad y=W(x),
\ee
\be   \label{3.3.4}
D_x^{(i)}=0 \qquad (i=1,2) \quad \mbox{as} \quad x\to \pm\infty.
\ee

Substituting equation (\ref{3.3.1}) in the equations that are obtained from
the third- and fourth-order approximation, we obtain
\be   \label{3.3.5}
(H+1)B_{xx}^{(1)} -C^{(1)}=0,
\ee
 and for $i=1,2$
\be   \label{3.3.6}
B_{xx}^{(i)}-C^{(i)}+\frac{1}{1-R}\left(
RB_{tt}^{(1)}-B_{tt}^{(2)}\right)=0.
\ee
%!!!!!!!!!!!!!!!!!!!!!
From equations (\ref{3.3.3}), (\ref{3.3.5}), and (\ref{3.3.6}) we get
\be \label{3.3.7}
\ba{l}
\Box_1 B^{(1)}=B_{tt}^{(2)},\\[2mm]
\Box_2 B^{(2)} =RB_{tt}^{(1)},
\ea
\ee
where $\Box_1$, $\Box_2$ are the dif\/ferential operators def\/ined by
\be \label{3.3.9}
\ba{l}
\ds
\Box_1 \equiv -H(1-R)\frac{\p^2}{\p x^2} +R\frac{\p^2}{\p t^2},\\[4mm]
\ds
\Box_2\equiv -(1-R)(1-W)\frac{\p^2}{\p x^2}+\frac{\p^2}{\p t^2}
+(1-R)\frac{\p W}{\p x} \frac{\p}{\p x}.
\ea
\ee
From equations (\ref{3.3.7})-(\ref{3.3.9}) we can get, after getting rid of
$B^{(1)}$ and substituting for
$W(x)$, the following dif\/ferential equation for the unknown function
$B^{(2)}(x,t)$
\be \label{3.3.11}
%\hspace*{-6pt}
\ba{l}
\ds
-H(1-R)(1-b-ax)B_{xxxx}^{(2)}+[H+R(1-b-ax)]B_{xxtt}^{(2)}\\[4mm]
\ds
\qquad -RaB_{xtt}^{(2)}+3H a(1-R)B_{xxx}^{(2)}=0
\ea
\ee
and for $f_4(x,t)$ we can get the following relation
\[
\ba{l}
\ds f_4(x,t) =\frac{1}{1-R} \left\{ R\left[ -\frac 12 B_{xxt}^{(1)}
+C_t^{(1)}+ D_t^{(1)}+\frac 12 \left(B_x^{(1)}\right)^2\right]
\right.
\\[4mm]
\ds \left. \phantom{f_4(x,t)=}+\frac 12 B_{xxt}^{(2)}
-C_t^{(2)}- D_t^{(2)}-\frac 12 \left(B_x^{(2)}\right)^2\right\}.
\ea
\]

\subsection{The f\/ifth- and sixth-order approximations}

Equations of the f\/ifth- and sixth-order approximation lead to, for
$i=1,2$
\be \label{3.4.1}
G_5^{(i)} =\frac{y^4}{24}
B_{xxxx}^{(i)}-\frac{y^3}{6}C_{xx}^{(i)}(x,t)-\frac{y^2}{2}
D_{xx}^{(i)}(x,t) + y E^{(i)}(x,t)+ F^{(i)}(x,t),
\ee
where $E^{(i)}(x,t)$ and $F^{(i)}(x,t)$ are arbitrary functions,
satisfy the folowing conditions:
\be \label{3.4.2}
E_x^{(i)}=0 \qquad (i=1,2), \quad \mbox{as} \quad x\to \pm \infty
\ee
and at $y=W(x)$
\be  \label{3.4.3}
E^{(2)}(x,t)=\left(-\frac{W^3}{3!} B_{xxx}^{(2)}+\frac{W^2}{2!}
C_x^{(2)}(x,t)+W D_x^{(2)}\right)_x,
\ee
\be  \label{3.4.4}
F_x^{(i)}=0 \qquad (i=1,2) \quad \mbox{as} \quad x\to \pm \infty.
\ee
Introducing equations (\ref{3.3.1})-(\ref{3.4.1})
in the boundary conditions, we
have the followind relations:
\be  \label{3.4.5}
\frac{(H+1)^3}{3!} B_{xxxx}^{(1)}-\frac{(H+1)^2}{2!} C_{xx}^{(1)}-
(H+1)D_{xx}^{(1)}+E^{(1)}=0
\ee
and for $i=1,2$
\be  \label{3.4.6}
\ba{l}
\ds \frac{1}{3!}
B_{xxxx}^{(i)}-\frac{1}{2!}C_{xx}^{(i)}-D_{xx}^{(i)}+ E^{(i)}+
\frac{1}{1-R}\Bigl[ \left(B_t^{(2)}-RB_t^{(1)}\right)B_{xx}^{(i)}
\\[3mm]
\ds \qquad +\left(B_{xt}^{(2)}-  RB_{xt}^{(1)}\right) B_x^{(i)}-
 \frac 12 B_{xxtt}^{(2)}+C_{tt}^{(2)} +D_{tt}^{(2)}\\[3mm]
\ds \qquad -R\left(-\frac 12 B_{xxtt}^{(1)}+C_{tt}^{(1)}+D_{tt}^{(1)}\right)+
B_x^{(2)}B_{xt}^{(2)}-RB_x^{(1)}B_{xt}^{(1)}\Bigr]=0.
\ea
\ee
Thus the problem is now reduced to solving equations (\ref{3.3.5}) and
(\ref{3.3.6}) for $B^{(i)}$ and $C^{(i)}$
and next equations (\ref{3.4.2}), (\ref{3.4.3}) and (\ref{3.4.6}) for
$D^{(i)}$ and $E^{(i)}$, where $i=1,2$.

\section{Case of progressive wave}

It must be remarked that our procedure is valid as long as
$a\gg \varepsilon^2$, otherwise a two-parameter analysis has to be carried out. Moreover, we shall
invoke the smallness of $a$ and write perturbation expansions for
$B^{(i)}$, $(i=1,2)$, in the form
\be  \label{4.1}
B^{(i)} =B_0^{(i)} +aB_1^{(i)} +a^2 B_2^{(i)}+\cdots
\ee
Substituting (\ref{4.1}) in (\ref{3.3.11}) and equating coef\/f\/icients of
$a^{(j)}$, $j=0,1,2,\ldots$ we get the
following system of dif\/ferential equations
\be \label{4.2}
\Box B_j^{(2)}=\Lambda B_{j-1}^{(2)} \qquad (j=0,1,2, \ldots), \quad
B_{-1}^{(2)}=0,
\ee
 where $\Box$, $\Lambda$
are two dif\/ferential operators def\/ined as
\[
\ba{l}
\ds \Box\equiv -H(1-R)(1-b)\frac{\p^4}{\p x^4}+[H+R(1-b)]\frac{\p^4}{\p
x^2 \p t^2},\\[3mm]
\ds \Lambda\equiv -xH(1-R)\frac{\p^4}{\p x^4}+xR\frac{\p^4}{\p x^2 \p
t^2}- 3H(1-R)\frac{\p^3}{\p x^3}+R\frac{\p^3}{\p x\p t^2}.
\ea
\]
Equation (\ref{4.2}), for
$j=0$, has the following general solution, for the
case of pure progressive waves,
\[
B_0^{(i)}=B_0^{(i)}(\xi)
\]
with
\[
\xi=x-\gamma t, \qquad \gamma^2 =\frac{H(1-b)(1-R)}{H+(1-b)R}.
\]

From equations (\ref{2.5}), (\ref{3.3.3}), and (\ref{4.1}) we get
\be \label{4.6}
C^{(2)}= \sum_{n=0}^\infty a^n \left[ b B_{n, xx}^{(2)}+
\left(xB_{n-1,x}^{(2)}\right)_x\right], \qquad B_{-1}^{(2)}=0.
\ee
Again substituting equations (\ref{4.1}), (\ref{4.6}) in
equation (\ref{3.3.6}) we get,
after equating coef\/f\/icients of $a^n$, $n=0,1,2,\ldots$.
\[
B_{0,x}^{(2)}=\lambda B_{0,x}^{(1)},
\qquad B_{1,x}^{(2)}=\frac{x}{1-b}B_{0,x}^{(2)}+\lambda B_{1,x}^{(1)},
\]
where
\[
\lambda = \frac{H}{1-b}.
\]

The elimination of $E^{(1)}$  in equations (\ref{3.4.5}) and
(\ref{3.4.6})
gives, for ``$a$'', the following system
of dif\/ferential equations
\be \label{4.10}
\ba{l}
\ds \left( H-\frac{\gamma^2 R}{1-R}\right) D_{\xi\xi}^{(1)}+
\frac{\gamma^2}{1-R} D_{\xi\xi}^{(2)}
=P_1B_{0,\xi\xi\xi\xi}^{(1)}+Q_1 B_{0,\xi}^{(1)} B_{0,\xi\xi}^{(1)},\\[4mm]
\ds
\frac{\gamma^2 R}{1-R} D_{\xi\xi}^{(1)}
+\left(1-b-\frac{\gamma^2}{1-R}\right) D_{\xi\xi}^{(2)}=
P_2B_{0,\xi\xi\xi\xi}^{(1)} +Q_2 B_{0,\xi}^{(1)} B_{0,\xi\xi}^{(1)},
\ea
\ee
where
\[
\ba{l}
\ds P_1=\frac{-H(2H^2+6H+3)}{6} +
\frac{\gamma^2[\lambda(1-2b)+R(2H+1)]}{2(1-R)},\\[3mm]
\ds P_2=\frac{\lambda}{6}(1-3b+2b^3) +\frac{\gamma^2}{2(1-R)}
[(2b-1)\lambda-R(2H+1)],
    \\[4mm]
\ds Q_1=\frac{\gamma}{1-R}(\lambda^2+2\lambda-3R),
\qquad
Q_2=\frac{\gamma}{1-R}[R(2\lambda +1)-3\lambda^2].
\ea
\]
For the non-trivial solution of
$D_{\xi\xi}^{(1)}$  and $D_{\xi\xi}^{(2)}$,
the following dif\/ferential equation for  $B_0^{(1)}$  should be satisf\/ied:
\be \label{4.16}
M_1 B_{0,\xi\xi\xi\xi}^{(1)}+M_2B_{0,\xi}^{(1)} B_{0,\xi\xi}^{(1)}=0,
\ee
where
\[
M_1=\left(1-b-\frac{\gamma^2}{1-R}\right) P_1-\frac{\gamma^2}{1-R}
P_2,
\quad M_2=\left(1-b-\frac{\gamma^2}{1-R}\right) Q_1-\frac{\gamma^2}{1-R}
Q_2.
\]
Def\/ine
\be \label{4.19}
\Gamma= B_{0,\xi}^{(1)}.
\ee

Thus equation (\ref{4.16}), by virtue of equation (\ref{4.19}), will be
transformed to the Boussinesq equation
\be \label{4.20}
M_1 \Gamma_{\xi\xi\xi}+M_2\Gamma \Gamma_\xi=0.
\ee

Helal \& Molines [3] mentioned that the general solution of equation
(\ref{4.20}) was found by Byrd and Friedmann [2] to be, in terms of the
Jacobi elliptic function $\mbox{sn}(u,k)$, as
\[
B_{0,\xi}^{(1)}=Y_1\left[ 1-\frac{3k^2}{k^2+1}\mbox{sn}^2(\delta\xi,
k^2)\right],
\]
where $Y_1$ is the greatest of the roots of the polynomial resulting
from integrating equation
(\ref{4.20}) twice and $k$ is the modulus of the Jacobean elliptic function,
and
\[
\delta =\frac 12 \left(- \frac{3AY_1}{k^2+1}\right)^{1/2}.
\]

For small values of $k$ the above elliptic function could be
calculated in terms of
trigonometric functions, see Milne-Thomson [11], thus we have
\be  \label{4.23}
\hspace*{-15.5pt}\ba{l}
\ds B_{0,\xi}^{(1)}= Y_1 \left\{ 1-\frac{3k^2}{k^2+1} \left[
\left(\frac 12 +\frac{k^2}{8} +\frac{k^4}{16}\right)+
\frac{k^4-64}{128} \cos 2\delta \xi-
\frac{8k^2+k^4}{64} \cos 4\delta \xi\right. \right.\\[4mm]
\ds %\phantom{B_{0,\xi}^{(1)}=}
\left. \left.
-\frac{k^4}{128}\cos 6\delta
\xi-   \delta \xi\left\{ \left(\frac{k^2}{2}
+\frac{k^4}{8}\right) \sin 2\delta \xi +
\frac{k^4}{16} \sin 4\delta \xi\right\} +\delta^2
\xi^2
\left\{ \frac{k^4}{8}+\frac{k^4}{8}\cos 2\delta \xi\right\}\right]
\right\}.
\ea
\ee

Substituting in equation (\ref{4.2}), for $B_{0,x}^{(2)}$ and
$B_{0,t}^{(2)}$, we get the following fourth-order linear partial
dif\/ferential equation
\be\label{4.24}
\ba{l}
\ds \Box B_1^{(2)} =\sum\limits_{n=1}^3
(A_n x\sin 2n\delta \xi+A_{n+6} \cos 2n\delta\xi)\\[4mm]
\ds \phantom{\Box B_1^{(2)} =}+\delta\xi
\sum\limits_{n=1}^2 (A_{n+3} x\cos 2n\delta\xi +A_{n+10}\sin 2n\delta
\xi)\\[4mm]
\ds \phantom{\Box B_1^{(2)} =} +\delta^2\xi^2(A_6 x\sin 2\delta \xi
+A_{13} \cos 2\delta \xi) +A_{10},
\ea
\ee
where the coef\/f\/icients $A_1, A_2,\ldots, A_{13}$ are
given at the end of the paper, as Appendix~1.

Solving equation (\ref{4.24}) for the unknown $B_1^{(2)}$, following Miller
[10], and calculating $B_{1,t}^{(2)}$  we get
\be \label{4.25}
\ba{l}
B_{1,t}^{(2)} =B_{0,t}^{(2)} +r_1 t^3 +r_2x^2 t+(r_3+r_4x^2+r_5
xt+r_6t^2) \sin 2\delta \xi \\[2mm]
\phantom{B_{1,t}^{(2)}=}+(r_7+r_8x^2 +r_9xt+r_{10}t^2) \sin 4\delta \xi +r_{11} \sin 6 \delta
\xi \\[2mm]
\phantom{B_{1,t}^{(2)}=}+(r_{12} +r_{13}x +r_{14} t+r_{15}x^3+ r_{16}x^2t +r_{17}xt^2+
r_{18}t^3) \cos 2\delta \xi\\[2mm]
\phantom{B_{1,t}^{(2)}=}+(r_{19}x+r_{20} t)\cos 4\delta \xi +(r_{21}x +r_{22} t) \cos 6\delta
\xi, \ea
\ee
where the coef\/f\/icients $r_1, r_2, \ldots, r_{22}$ are also given at
the end of the paper, as Appendix~2.

Taking into consideration the value of
$B_{0,x}^{(1)}$  from  equation (\ref{4.23}), we can get
$B_{0,x}^{(2)}$ and thus, using (\ref{4.25}) for
$B_{1,t}^{(2)}$ we can get $B_{1,t}^{(1)}$
\[
B_{1,t}^{(1)}= \frac{1}{\lambda}\left( B_{1,t}^{(2)}
-\frac{x}{1-a}B_{0,t}^{(2)} \right).
\]

In order to account for the nonlinear ef\/fects the $O(\varepsilon^4)$
equations have to be considered as well. Thus bearing in mind the
linear system of equations (28), the principal and
secondary determinants of this system, we come to the result that
\[
D_t^{(i)}=0, \qquad (i=1,2).
\]
Hence $f_4(x,t)$  may be rewritten in the simplif\/ied form
\be \label{4.28}
\hspace*{-17.62pt}\ba{l}
\ds f_4(x,t) =\frac{1}{2(1-R)}\left\{ \left(
(\lambda-R)+2(1+H)R-2\lambda(ax+b)+
\frac{ax\lambda(1-2b)}{1-b}\right) B_{0,xxt}^{(1)}\right.\\[4mm]
\ds \qquad +(R(2H+1)+\lambda(1-2b))B_{1,xxt}^{(1)}+
\left(2(R-\lambda^2)-
\frac{2ax\lambda^2(ax+2)}{1-b}\right)(B_{0,x}^{(1)})^2\\[4mm]
\ds \qquad \left. +a^2(R-a)(B_{1,x}^{(1)})^2+
\frac{2a\lambda b}{b-1} B_{0,xt}^{(1)}+2a\left(R-\lambda^2-
\frac{ax\lambda^2}{1-b}\right)B_{0,x}^{(1)} B_{1,x}^{(1)}\right\}.
\ea
\ee
Hence $f(x,t)$ will take the form
\[
f(x,t) =1+\varepsilon^2 \left\{ \frac{(R-\lambda)(b-1)+\lambda b
x}{(1-R)(b-1)} B_{0,t}^{(1)} +
\frac{a(R-\lambda)}{1-R}B_{1,t}^{(1)}\right\} +\varepsilon^4
f_4(x,t)+O(\varepsilon^6),
\]
where $f_4(x,t)$  is given by (\ref{4.28}) and $B_{0,t}^{(1)}$
and $B_{1,t}^{(1)}$ are given by (\ref{4.23}) and (\ref{4.25})
respectively.

\section{Presentation of results and discussion}
The number of terms which has been obtained seems to be a good
measure for the purpose of illustrating the ef\/fect of the parameters
the density ratio, $R$, the thickness
ratio, $H$, and the obstacle height, $L$. The error, dif\/ference
between the fourth and second order approximations, in the
interfacial prof\/ile for the two approximations is of order
$10^{-6}$.
Thus we limit our calculations up to the second-order approximation,
as well as we considered the following values for the description of
the triangular obstacle: $x_m=13$  and  $x_e=30$.

We studied the ef\/fect of the density  ratio, $R$,
on the wave prof\/iles at the interfacial
surface. Three values of R have been considered, namely $R=0.7$,
$0.8$, and $0.9$ for f\/ixed values of $H$, $L$, and $t$. It is clear
that as $R$ decreases, there
is a kind of violent oscillations in the obstacle region. This
phenomena vanishes gradually as ``$R$''
increases. An important remark must be mentioned is that, for the
interfacial wave in the downstream region, the period of oscillation
is much longer for the case when the two f\/luids
are of very nearly equal density than that of signif\/icant dif\/ferent densities.
This is due to the fact that the presence of the upper f\/luid has the
ef\/fect of decreasing the velocity of propagation of the wave which
consequently causes the decrease of the potential energy
of a given deformation of the interface as well as the increase of
the inertia. This result comes in good agreement with Lamb [6], who gave
a marvelous natural example for such a phenomena, occurring near
the mouths of some of the Norwegian f\/iord,
when there is a layer of fresh water over salt water.

The interfacial wave prof\/iles, $f(x,t)$, has been studied for dif\/ferent
values of the thickness ratio, $H$, namely $H = 0.3$, $0.5$, and $0.6$
while the other parameters $R$, $L$, and $t$ are f\/ixed.
It is clear that as $H$ increases, there is an increase in
the amplitude of the wave along the obstacle interval, as well
as an increase in the wave length.

We study the ef\/fect of changing the triangle height, $L$. Three
values of $L$ have been considered, namely $L = 0.1$, $0.2$, and
$0.25$ for f\/ixed values of $R$, $H$, and $t$. For the
interfacial wave, as $L$ increases a kind of violent disturbance in
the wave prof\/ile appears, starting by a sudden increase in the
prof\/ile, ending by a steep decrease at the beginning of the
downstream interval. The behaviour of that solution can be
interpreted, following Kakutani [4], as follows: a given smooth
waveform will propagate along the characteristic curves, gradually
steepen its shape due to nonlinear interactions,
and then the dispersive term will begin to play its role to
balance this steeping.

\section*{Appendix 1}
\[
A_1= W_1\left(-4+6k^2+ \frac{1}{16}k^4\right) \qquad
A_2=W_1(2k^4-8k^2) \qquad A_3=W_1\left(-\frac{27}{16}k^4\right)
\]
\[
A_4=W_1(4k^2-2k^4)\qquad A_5=4W_1k^4 \qquad A_6=W_1k^4
\]
where
\[
W_1=[\gamma^2-H(1-R)]\left(-\frac{3Y_1k^2\delta^3}{k^2+1}\right)
\]
and
\[
A_7=W_2\left(2-2k^2-\frac{9}{32}k^4\right) \qquad
A_8 =W_2\left(2k^2- \frac 14 k^4\right) \qquad
A_9=\frac{9}{32}W_2 k^4
\]
\[
A_{10}= \frac 14 W_2k^4 \qquad
A_{11}=W_2\left(2k^2-\frac 12 k^4\right) \qquad
A_{12}=W_2k^4 \qquad
A_{13}=-\frac 12 W_2k^4
\]
where
\[
W_2=[\gamma^2-3H(1-R)]\left(-\frac{3Y_1k^2\delta^2}{k^2+1}\right)
\]
and
\[
A_{14}=H(1-R)(b-1) \qquad
A_{15}=H+1-b \qquad
A_{16}= \frac{1}{4\gamma \delta}(2A_1-A_4-A_6)
\]
\[
A_{17}=\frac{1}{2\gamma}\delta A_{12}\qquad
A_{18}=\frac{1}{2\gamma}(A_{11}-A_{13})\qquad
A_{19}=\frac{1}{16\gamma\delta}(4A_2-A_5)
\]
\[
A_{20}=\frac{1}{4\gamma}A_{12} \qquad
A_{21}= \frac{1}{6\gamma\delta}A_3\qquad
A_{22}=\frac{1}{4\gamma\delta}(A_{13}-2A_7-A_{11})
\]
\[
A_{23}=-\frac{1}{2\gamma}(A_4+A_6)\qquad
A_{24}=-\frac{1}{2\gamma}\delta A_{13}\qquad
A_{25}=-\frac{1}{16\gamma\delta}(4A_8+A_{12})
\]
\[
A_{26}=-\frac{1}{4\gamma}A_5 \qquad
A_{27}=-\frac{1}{6\gamma\delta}A_9\qquad
A_{28}=\frac{1}{A_{15}-2\gamma^2} \qquad
A_{29}=-\frac{2}{\gamma}A_{14}A_{28}
\]
\[
A_{30}=-2A_{28}\qquad
A_{31}=A_{15}A_{28}\qquad
A_{32}=-\frac{1}{\gamma}A_{15}A_{28}\qquad
A_{33}=\gamma A_{15}A_{28}
\]
\[
A_{34}=\frac{1}{2\gamma}A_{14}A_{28}\qquad
A_{35}=3\gamma A_{28}\qquad
A_{36}= \frac{1}{2\gamma}A_{15}A_{28}\qquad
A_{37}=\frac{1}{2\gamma}A_{28}
\]
\[
A_{38}=(\gamma^2A_{15}+2A_{14})A_{28}\qquad
A_{39}=\frac{1}{2\gamma}(6A_{14}+\gamma^2A_{15})A_{28}\qquad
A_{40}=2A_{15}A_{28}
\]
\[
A_{41}=\frac{1}{4\gamma\delta}(2A_1-A_4-A_6+2\delta[A_{11}-A_{13}])
\qquad A_{42}=-\delta A_6 \qquad
A_{43}=\frac 12 \gamma\delta A_6
\]
\[
A_{44}=\frac 12(A_{13}-A_{11})\qquad
A_{45}=\frac{1}{16\gamma\delta}(4A_2-A_5+4\delta A_{12})\qquad
A_{46}=-\frac 14 A_{12}
\]
\[
A_{47}=-\frac{1}{2\gamma}(A_4+A_6+\delta A_{13})\qquad
A_{48}=\frac 12(A_4+A_6+2\delta A_{13})\qquad
A_{49}=-\frac 12 \gamma \delta A_{13}
\]
\[
A_{50}=\frac 14 A_5\quad \ \
A_{51}=3A_{38}^2 A_{35}\quad \ \
A_{52}=3A_{38}^2 A_{40}\quad  \ \
A_{53}=6A_{38}(A_{30}A_{39}+A_{32}A_{40})
\]
\[
A_{54}=6A_{38}A_{30}A_{35}\qquad
A_{55}=6A_{38}(A_{30}A_{40}+A_{32}A_{35})\qquad
A_{56}=-3A^2_{35}A_{38}
\]
\[
A_{57}=-3A_{38}(A_{40}^2+2A_{39}A_{35})
\qquad
A_{58}=2A_{35}A_{40}A_{38}
\]
\[
A_{59}=A_{29}+2A_{31}A_{38}-2A_{39}A_{40}+3A_{38}^2A_{32}+A_{57}
\qquad
A_{60}=A_{30}-A_{35}^2
\]
\[
A_{61}=A_{31}+2A_{38}A_{32}-A_{40}^2-2A_{39}A_{35}+3A_{38}^2A_{30}+A_{58}
\]
\[
A_{62}=A_{32}+2A_{38}A_{30}-2A_{35}A_{40}+A_{56}
\]
\[
A_{63}=2(A_{38}A_{36}+A_{32}A_{40}+A_{32}A_{39})-A_{40}-6A_{39}A_{35}A_{40}
+A_{53}
\]
\[
A_{64}=A_{36}+2(A_{31}A_{35}+A_{32}A_{40})-3A_{35}(A_{35}A_{39}+A_{40}^2)
+A_{55}
\]
\[
A_{65}=2(A_{30}A_{40}+A_{32}A_{35})+A_{54}-3A_{35}^2A_{40}\qquad
A_{66}=A_{39}+2A_{38}A_{40}+A_{51}
\]
\[
A_{67}=A_{40}+2A_{38}A_{35}\qquad
A_{68}= 2A_{38}A_{39}+A_{52}
\]
\[
A_{69}= A_{25}+\frac{1}{16\delta^2}(2A_{61}A_{26}+A_{62}A_{50})
\qquad
A_{70}= A_{50}+2A_{38}A_{26}
\]
\[
A_{71}=\frac 12 A_{38}(A_{50}+2A_{38}A_{26})\qquad
A_{72}=\frac{1}{4\delta}(A_{35}A_{50}+2A_{26}A_{67})
\]
\[
A_{73}=\frac{1}{4\delta}(2A_{66}A_{26}+A_{67}A_{50})
\qquad
A_{74}=A_{22}+\frac{1}{2\delta^2}(A_{60}A_{49}+A_{61}A_{47}+2A_{62}A_{48})
\]
\[
A_{75}=A_{48}+2A_{38}A_{47}\qquad
A_{76}=A_{49}+\frac 12 A_{38}(A_{48}+2A_{38}A_{47})
\]
\[
A_{77}=\frac{1}{2\delta}(A_{35}A_{48}+2A_{67}A_{47})\qquad
A_{78}=\frac{1}{\delta}(A_{35}A_{49}+2A_{67}A_{48}+A_{66}A_{47})
\]
\[
A_{79}=\frac{1}{4\delta}(A_{35}A_{46}+A_{67}A_{45})\qquad
A_{80}=A_{41}+\frac{1}{2\delta^2}(A_{60}A_{43}+3A_{61}A_{17}+A_{62}A_{42})
\]
\[
A_{81}=A_{42}+3A_{38}A_{17}\qquad
A_{82}=A_{43}+A_{38}A_{42}+3A_{38}^2A_{17}
\]
\[
A_{83}=A_{44}+A_{38}A_{41}+\frac{1}{2\delta^2}(A_{61}A_{42}+A_{62}A_{43})
\]
\[
A_{84}=\frac 13(A_{38}A_{43}+A_{38}^2A_{42})+A_{38}^3A_{17}
\]
\[
A_{85}=\frac{1}{4\delta^3}(3A_{63}A_{17}+A_{64}A_{42}+A_{65}A_{43})+
\frac{1}{2\delta}(A_{35}A_{44}+A_{67}A_{41})
\]
\[
A_{86}=\frac{1}{\delta}(3A_{66}A_{17}+A_{35}A_{43}+A_{67}A_{42})
\qquad
A_{87}=\frac{1}{2\delta}(A_{66}A_{42}+A_{67}A_{43}+3A_{68}A_{17})
\]
\[
A_{88}=\frac{1}{2\delta}(A_{35}A_{42}+3A_{67}A_{17})
\]

\section*{Appendix 2}
\[
r_1=-\frac 16 A_{10}\qquad
r_2=\frac{1}{8\delta^3}A_{37}[2\gamma \delta(A_{74}-A_{85})-A_{78}-A_{83}]
\]
\[
r_3=\frac{1}{8\delta^3}A_{37}[2\gamma\delta(A_{47}-A_{88})-A_{81}]
\qquad
r_4=\frac{1}{4\delta^3}A_{37}[\gamma\delta(A_{75}-A_{86})-A_{82}]
\]
\[
r_5=\frac{1}{8\delta^3}A_{37}[2\gamma\delta(A_{76}-A_{87})-3A_{84}]
\]
\[
r_6=\frac{1}{64\delta^3}A_{37}[4\gamma\delta(A_{69}-A_{79})-A_{73}-A_{46}
-A_{38}A_{45}]
\qquad
r_7=\frac{1}{16\delta^2}\gamma A_{37}A_{26}
\]
\[
r_8=\frac{1}{16\delta^2}\gamma A_{37}A_{70}
\qquad
r_9=\frac{1}{16\delta^2}\gamma A_{37}A_{71}
\]
\[
r_{10}=\frac{1}{216\delta^3}A_{37}[\gamma(6\delta A_{27}-A_{67})-A_{38}A_{21}]
\qquad
r_{11}=\frac{1}{4\delta^2}A_{37}A_{80}
\]
\[
r_{12}=\frac{1}{8\delta^3}A_{37}(2\gamma\delta A_{77}+A_{75}-A_{86})
\qquad
r_{13}=\frac{1}{4\delta^3}A_{37}[\gamma\delta(A_{78}+A_{83})+A_{76}-A_{87}]
\]
\[
r_{14}=\frac{1}{4\delta^2}\gamma A_{37}A_{88}
\qquad
r_{15}=\frac{1}{4\delta^2}\gamma A_{37}A_{81}
\qquad
r_{16}=\frac{1}{4\delta^2}\gamma A_{37}A_{82}
\]
\[
r_{17}=\frac{1}{4\delta^2}\gamma A_{37}A_{84}
\qquad
r_{18}=\frac{1}{64\delta^3}A_{37}[A_{70}+4\gamma \delta(A_{72}+A_{45})]
\]
\[
r_{19}=\frac{1}{32\delta^3}A_{37}[A_{71}+2\gamma\delta(A_{73}+A_{46}+
A_{38}A_{45})]\qquad
r_{20}=\frac{1}{36\delta^2}\gamma A_{37}A_{21}
\]
\[
r_{21}=\frac{1}{36\delta^2}\gamma A_{37}A_{38}A_{21}
\qquad
r_{22}=\frac{1}{36\delta^2}\gamma A_{38}A_{21}A_{43}
\]

 \label{malek-lp}
\end{document}